\documentclass[aps,prl,twocolumn,unsortedaddress,floatfix,nofootinbib,superscriptaddress]{revtex4-2}
\usepackage{amsthm}
\usepackage{amsfonts}
\usepackage{siunitx}
\usepackage{amsmath}
\usepackage{amssymb}
\usepackage{graphicx}
\usepackage{verbatim}
\usepackage[colorlinks]{hyperref}
\usepackage{tikz}
\usepackage{braket}
\usepackage{xcolor}
\setcitestyle{super,open={},close={}}
\renewcommand{\onlinecite}[1]{\nocite{#1}\citenum{#1}}

\makeatletter
\def\@ssect@ltx#1#2#3#4#5#6[#7]#8{%
  \def\H@svsec{\phantomsection}%
  \@tempskipa #5\relax
  \@ifdim{\@tempskipa>\z@}{%
    \begingroup
      \interlinepenalty \@M
      #6{%
       \@ifundefined{@hangfroms@#1}{\@hang@froms}{\csname @hangfroms@#1\endcsname}%
       {\hskip#3\relax\H@svsec}{#8}%
      }%
      \@@par
    \endgroup
    \@ifundefined{#1smark}{\@gobble}{\csname #1smark\endcsname}{#7}%
  }{%
    \def\@svsechd{%
      #6{%
       \@ifundefined{@runin@tos@#1}{\@runin@tos}{\csname @runin@tos@#1\endcsname}%
       {\hskip#3\relax\H@svsec}{#8}%
      }%
      \@ifundefined{#1smark}{\@gobble}{\csname #1smark\endcsname}{#7}%
      \addcontentsline{toc}{#1}{\protect\numberline{}#8}%
    }%
  }%
  \@xsect{#5}%
}%
\makeatother

\definecolor{linkcolor}{RGB}{0,83,166}
\hypersetup{
  colorlinks = true,
  allcolors = {linkcolor}
}

\begin{document}
\newcommand{\mytitle}{Quantum dynamics in frustrated Ising fullerenes}
\title{\mytitle}

\newcommand{\affildw}{D-Wave Quantum Inc., Burnaby, British Columbia, Canada}

\author{Alejandro Lopez-Bezanilla}
\email[]{alejandrolb@gmail.com}
\affiliation{Theoretical Division, Los Alamos National Laboratory, Los Alamos, NM 87545, USA}

\author{William Bernoudy}
\affiliation{\affildw}

\author{Kelly Boothby}
\affiliation{\affildw}

\author{Jack Raymond}
\affiliation{\affildw}

\author{Alberto Nocera}
\affiliation{Department of Physics and Astronomy and Quantum Matter Institute, University of British Columbia, Vancouver, British Columbia, Canada}

\author{Andrew D.~King}
\email[]{aking@dwavesys.com}
\affiliation{\affildw}

\date{\today}
\begin{abstract}
The complex energy landscapes exhibited by frustrated magnetic systems undergoing quantum fluctuations are a challenge to accurately simulate, 
and thus of great interest for testing diverse qubit platforms in the field of quantum simulation. 
This study experimentally demonstrates quantum fluctuations lifting the degenerate ground-state manifold of 
classical magnetic configurations in fullerene Ising models with resonating dimers. 
The interplay between degeneracy and quantum fluctuations makes these boundary-free models a suitable benchmark for quantum simulators. 
Indeed, we observe significant performance improvement across generations of superconducting quantum annealers, 
showing the potential of highly symmetric, frustrated systems for assessing the precision of quantum-simulation technologies.
\end{abstract}

\maketitle

\def\title#1{\gdef\@title{#1}\gdef\THETITLE{#1}}

\section{Introduction}
 
Magnetic frustration arises from the inability of magnetic moments to simultaneously satisfy all energetic terms; this can result in highly degenerate ground-state manifolds with extensive ground-state entropy\cite{vannimenus_theory_1977,wang_artificial_2006}. Quantum fluctuations partially resolve this degeneracy by allowing the magnetic system to explore superpositions of classically degenerate states through mechanisms like tunneling\cite{shannon_quantum_2012}. Quantum terms, such as a transverse field applied to an Ising model, mix the degenerate configurations and effectively select states that minimize the overall energy\cite{gardner_magnetic_2010}.
Known as order-by-disorder\cite{villain_order_1980,hanai_nonreciprocal_2024,schumm_primary_2024,champion_er_2003,zhitomirsky_quantum_2012}, these perturbative corrections introduce effective interactions that, lifting degeneracy by favoring configurations that maximize quantum coherence and stability, can lead to the emergence of non-Ising universal behavior~\cite{moessner_ising_2001,savary_quantum_2017}. These geometrically frustrated systems, with their wide range of phases and phenomena, are especially appealing as benchmarks for quantum simulation.

Simulating frustrated magnetic systems requires precise control over the Hamiltonian, as phenomena such as order-by-disorder depend on the accurate realization of the degenerate classical ground-state manifold to serve as the foundation for quantum effects. In recent years, quantum experiments have been conducted in Ising-like systems manufactured from controllable building blocks, including superconducting qubits \cite{barends_digitized_2016,king_observation_2018,ali_quantum_2024,miessen_benchmarking_2024,andersen_thermalization_2025}, neutral atoms \cite{keesling_quantum_2019,ebadi_quantum_2021,scholl_quantum_2021}, and trapped ions \cite{kim_quantum_2010}. Together, these platforms represent a complementary toolkit, offering diverse approaches for simulating quantum effects. To benchmark these various methods of quantum simulation, this paper proposes a series of boundary-free frustrated lattices and demonstrates their use with quantum annealers based on superconducting qubits.

\begin{figure}
  %\hrule
  \includegraphics{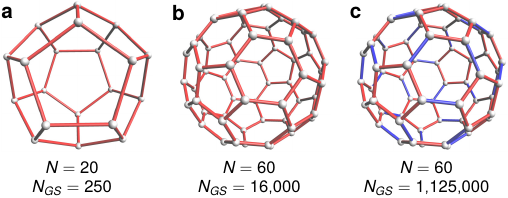}
  %\hrule
  \caption{
  {\bf Geometrically-frustrated Ising boundary-free systems.} 
    Representation of the antiferromagnetic 20-node dodecahedron, {\bf a} and 60-node fullerene {\bf b}, and {\bf c}. The fullerene's blue and red edges in {\bf c} form two symmetry classes that can be assigned ferromagnetic and antiferromagnetic exchange respectively, leading to greater frustration and higher ground-state degeneracy, $N_{\text GS}$, than in the purely antiferromagnetic case {\bf b}. }
  \label{fig:1}
\end{figure}

For a series of boundary-free frustrated magnetic lattices, embedded in D-Wave quantum annealers, this paper demonstrates the partitioning by quantum fluctuations of the ground-state manifold of classical magnetic configurations. By facilitating quantum tunneling and local resonance processes within magnetic dimers, in an environment free of thermal fluctuations,  we observe that energetically equivalent classical spin configurations paired into singlet bonds experience energy reduction depending on their local environment. These systems offer a hardware-efficient and versatile cross-platform benchmark that can potentially be evaluated in superconducting quantum annealers, neutral atom arrays~\cite{barredo_synthetic_2018}, and trapped-ion systems.

All systems have uniform-magnitude couplings $J_{ij} = \pm 1$ on the bonds $ij$ of a three-regular fullerene graph (Fig.~\ref{fig:1}); due to the odd connectivity, any two isoenergetic classical states must have Hamming distance $\geq 2$.  As such, a perturbative transverse field lifts the classical ground-state degeneracy with a second-order tunneling term arising from resonating bonds. Accurately reproducing the density of such bonds in out-of-equilibrium quantum dynamics is a highly sensitive task for quantum simulation, which we demonstrate here.

\section{Frustrated Ising Fullerenes}

We simulate the lattices displayed in Figure~\ref{fig:1}. Nearest-neighbor couplings $J_{ij}$ are defined on a fullerene structure~\cite{balaban_graph_1995,tomanek_guide_2014,prahlaadh_quantum_2022}, a closed mesh in which sites have valency three and elementary plaquettes have five or six sites. Specifically we study the dodecahedron ($N=20$) and Buckminsterfullerene ($N=60$; buckyball) structures.

The dodecahedron (Figure~\ref{fig:1}a) consists of 20 qubits interconnected in a pentagonal topology, and exhibits 250 degenerate ground states in a nearest-neighbor all-AFM coupling.  The odd cycles in the fullerene graph shown in Figure~\ref{fig:1}b ensure magnetic frustration, leading to $\sim$$10^4$ classically degenerate ground states when enforcing an all-AFM coupling between neighboring qubits of the Ising model.
In Figure~\ref{fig:1}c, two types of couplers are introduced in the system: red couplers connect qubits $i$ and $j$ in AFM interactions with $J_{ij}=+1$ while blue couplers establish a ferromagnetic (FM) interaction between two qubits with $J_{ij}=-1$. Red couplers form the five-site cycles, whereas blue couplers interconnect qubits in distinct five-site cycles; each qubit is incident to two red couplers and one blue coupler. 
Notably, couplers can be classified according to their automorphism class, indicating the symmetry of the system: the graph representing the spins is vertex transitive, meaning that any spin can mapped to any other spin by a graph automorphism.  This combination of FM and AFM interactions renders the lattice ``fully frustrated'', with both pentagonal and hexagonal plaquettes frustrated, resulting in over a million ground states. 

The transverse-field Ising model (TFIM) Hamiltonian governing the quantum dynamics of these fullerenes is given by
\begin{align}
  \mathcal H &= \Gamma \mathcal H_D + \mathcal J\mathcal H_I\\
  \mathcal H_D &= -\sum_i{\sigma_i^x}\\
  \mathcal H_I &= \sum_{<i,j>}{J_{ij}\sigma_i^z\sigma_j^z},
\end{align}
where $\sigma_i^x$ and $\sigma_i^z$ are Pauli operators; $\Gamma$ and $\mathcal J$ are the transverse field and the classical Ising energy scale, which respectively weight the driver Hamiltonian $\mathcal H_D$ and the Ising Hamiltonian $\mathcal H_I$. These systems can be embedded directly as subgraphs of the qubit connectivity graphs of the quantum computers used in this study. 

When $\Gamma{=}0$ and $\mathcal J{>}0$, these three Ising models have respective classical ground-state degeneracies $N_{\text{GS}}=250$, $16{,}000$, and $1{,}125{,}000$. Upon the addition of a perturbative transverse field, say $\Gamma{=}\epsilon{\rightarrow}0$ and $\mathcal J{=}1$, quantum fluctuations lift the degeneracy of the classical ground-state manifold, and the quantum ground state becomes a superposition of classical ground states.  The amplitudes of this perturbative ground state do not reflect the ``fair sampling'' of the $T=0$ Boltzmann distribution\cite{dickson_thermally_2013,lanting_experimental_2017,mandra_exponentially_2017,zhang_advantages_2017}, but rather a distinct and delicate quantum state that we propose here as a benchmark for quantum simulation.

\begin{figure}
    \centering
    \includegraphics[width = 0.995\linewidth]{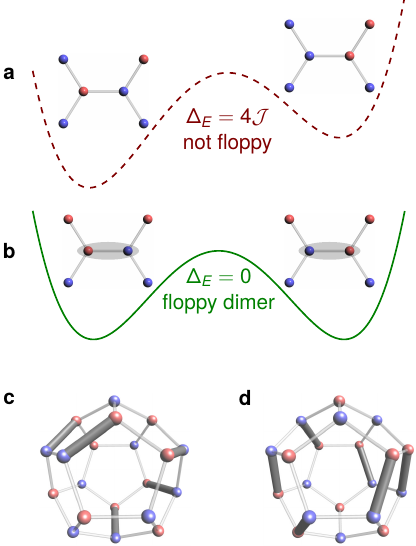}
 \caption{
 {\bf Floppy dimers and resonance in an AFM dodecahedron.} 
 Here, all couplers are antiferromagnetic with $J_{ij}=1$.  A floppy dimer is formed by pairing two spins with opposite orientations (blue and red) in a manner that facilitates consistent observation of the same type of nearest neighbor spins through transverse-field mediated tunneling within the dimer (exchange of spin orientations).  A regular dimer in {\bf a} tunnels between energy levels split by a gap $\Delta_E=4J$, while a ``floppy dimer'' in {\bf b} (thick gray line) tunnels between degenerate classical states.  Both {\bf c} and {\bf d} depict classical ground states, with floppy dimers indicated as thick couplings.  Their overlaps with $\psi_e$, $|\!\braket{\psi|\psi_\epsilon}\!|^2$, are determined by the structure of how floppy dimers connect the classical ground-state manifold.  For {\bf c}, $|\!\braket{\psi|\psi_\epsilon}\!|^2=0.0034$ is below average among the 250 classical ground states, whereas the value for {\bf d}, $0.0063$, is above average.  These represent two of the five symmetry classes of classical ground states under site relabeling.} \label{fig:2}
 \end{figure}

Any two classical ground states with Hamming distance $=2$ differ by a ``floppy dimer'', meaning that two coupled spins can cotunnel between the two configurations upon the application of quantum fluctuations.  A non-floppy dimer and a floppy dimer are depicted in Fig.~\ref{fig:2}a and b.  Fig.~\ref{fig:2}c and d show two distinct classical ground states of a dodecahedron, with floppy dimers indicated as thick gray lines.

\begin{figure}
  %\hrule
  \includegraphics[scale=.8]{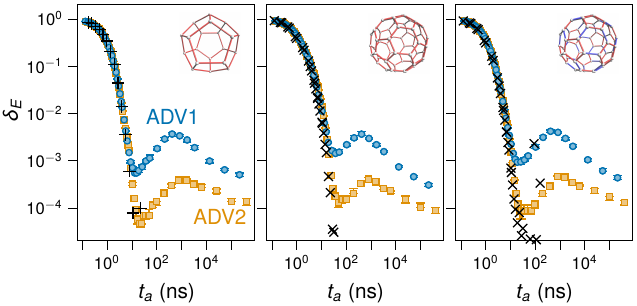}
  %\hrule
  \caption{{\bf Residual energy density.}  As the system approaches the classical ground-state manifold, residual energy density $\delta_E$ decreases sharply from $1$ as $t_a$ increases from the fast-anneal limit.  Classical simulations values are shown in black: ($+$: exact diagonalization for $N=20$; $\times$: MPS for $N=60$).  They show good agreement with observed QA dynamics up to the point where environmental effects arise.  The lower-noise ADV2 QPU drives to significantly lower energies than ADV1.  For longer anneals ($t_a>\SI{10}{ns}$), large step sizes lead to inconsistent energies in MPS.}\label{fig:3}
\end{figure}

We can consider a transverse field in the perturbative limit $\Gamma=\epsilon\rightarrow 0$.  The ground state $\ket{\psi_\epsilon}$ of $\epsilon\mathcal H_D+\mathcal H_I$ is a superposition of classical ground states, given by the principal eigenvector of the tunneling matrix among the classical ground states: rows and columns correspond to classical ground states and an entry corresponding to two states differing by a dimer flip is $-1$, or zero otherwise.  We can compare $\ket{\psi_\epsilon}$ with the uniform symmetric superposition $\ket{\psi_0}$ of classical ground states.  This has already been done in the case of single-spin tunnneling in small Villain models\cite{matsuda_groundstate_2009}.

We simulate the nonequilibrium approach toward $\ket{\psi_\epsilon}$ using quantum processing units (QPUs) from two generations of D-Wave annealing quantum computers: an Advantage\texttrademark\ system and an Advantage2\texttrademark\ prototype; we denote these ADV1 and ADV2 respectively as in Ref.~\onlinecite{king_beyondclassical_2025}.  They realize time-dependent transverse-field Ising Hamiltonians
\begin{equation}
\mathcal H(s) = \Gamma(s)\mathcal H_D + \mathcal J(s)\mathcal H_I, 
  \end{equation}
  where $\Gamma(s)$ and $\mathcal J(s)$ are the transverse and Ising energy scales at normalized time $s=t/t_a$ given by each processor's annealing schedule.  In the fast-quench limit $t_a\rightarrow 0$, the action of the Hamiltonian becomes nil and the output reflects the initial state, which is the ground state of $\mathcal H_D$:  equal superposition of computational basis states.  In other words, the final quenched state at $t=t_a$ should be a uniform distribution of all classical states (random, uncorrelated output).  In the slow-quench limit $t_a\rightarrow \infty$, an ideal closed quantum system will adiabatically approach $\ket{\psi_\epsilon}$ (we emphasize that the $s$-dependent ground state is discontinuous at $s{=}1$, and $\ket{\psi_\epsilon}$ is the ground state of the system in the limit $\Gamma\rightarrow 0$; the ideal system spends no time at $\Gamma=0$ and therefore will not discontinuously jump to the ground state at $\Gamma=0$, i.e., $\ket{\psi_0}$).  However, due to the limited coherence and precision of experimental platforms, we expect a thermalizing system to explore a broad space of low-energy classical states, deviating from the statistics of $\ket{\psi_\epsilon}$ toward $\ket{\psi_0}$.  This deviation provides a benchmark for quantum simulators by measuring a platform’s ability to maintain fidelity to the target quantum state, which requires precise control and advanced hardware.

\section{Results}

\begin{figure*}
  %\hrule
  \includegraphics[scale=1]{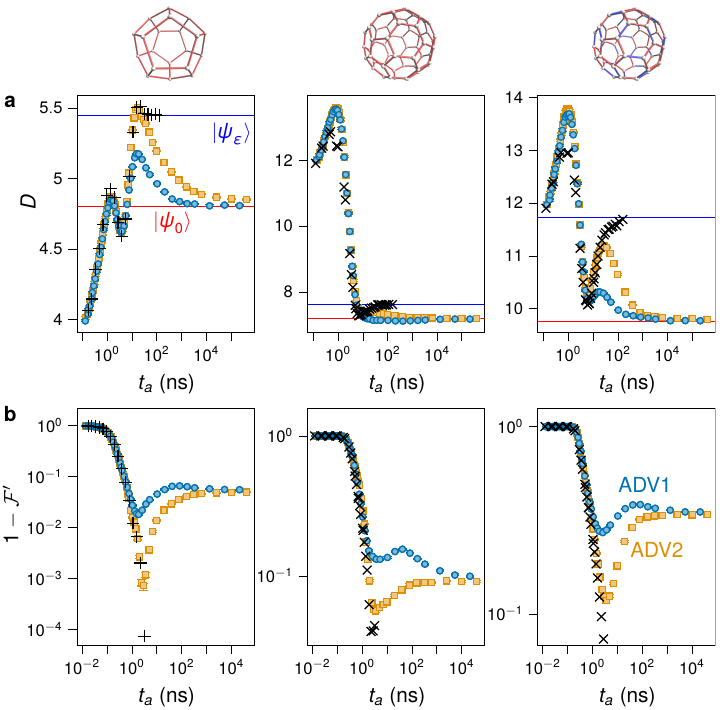}
  %\hrule
  \caption{{\bf Floppy dimers and fidelity.} {\bf a}, In an accurate simulation, the average number of floppy dimers, $D$, should reflect the perturbative ground state $\ket{\psi_\epsilon}$ in the large-$t_a$ limit.  In a thermalized quantum simulation, we expect the statistics of sampling from a uniform superposition over classical ground states $\ket{\psi_0}$.  For fast anneals, we observe nonmonotonicity in $D$ arising from competition between excited and ground states; this is reflected in both quantum and classical experiments.  {\bf b}, The binned fidelity $\mathcal F'$, calculated here over $10^5$ samples, quantifies adherence to the perturbative ground state $\ket{\psi_\epsilon}$, and clearly differentiates simulation quality between ADV1 and ADV2.  Infidelity for $10^5$ fair samples (95\% confidence interval) is below the axis limits.}\label{fig:4}
\end{figure*}

Along with quantum annealing (QA), we use exact diagonalization (ED) and matrix-product state (MPS) methods to simulate the model dynamics of the time-dependent QA Hamiltonian for varying $t_a$.  Although the QPUs' fastest accessible $t_a$ is $\SI{5}{ns}$, we can simulate faster quenches by reducing the magnitude of $J_{ij}$.  We plot all data using a calibrated equivalent time, assuming $|J_{ij}|=1$: all such $t_a$ are given relative to the ADV1 annealing schedule as in Ref.~\onlinecite{king_beyondclassical_2025} with physical annealing times for ADV2 roughly 1.75 times faster; details of the fidelity-based time calibration are given in the Appendix. MPS methods are as in Ref.~\onlinecite{king_beyondclassical_2025}; QA methods use public API settings available to all users.

  Fig.~\ref{fig:3} shows the residual energy density, defined by:
  \begin{equation}
  \delta_E = 1- \frac{\langle \mathcal H_I\rangle}{E_0}  
  \end{equation}
where $E_0 < 0$ is the ground-state energy of the classical Ising Hamiltonian $\mathcal H_I$ at the end of the anneal.  These observables are in good agreement between classical (ED and MPS) and quantum (QPU) experiments for short annealing times, but effects of QPU decoherence are apparent for $t_a > \SI{10}{ns}$, consistent with similar observations in analog QA \cite{king_coherent_2022}, digitized QA~\cite{miessen_benchmarking_2024} and digital-analog QA~\cite{andersen_thermalization_2025}.  ADV2, which has lower noise and a higher energy scale than ADV1, achieves lower energies both before and after the onset of thermal excitations.

  Fig.~\ref{fig:4}a shows $D$, the average number of floppy dimers in the classical output states, which can be defined as an operator
  \begin{equation}
      D = \sum_{<i,j>}(1-J_{i,i'}J_{i,i''}\sigma_{i'}^z\sigma_{i''}^z)(1-J_{j,j'}J_{j,j''}\sigma_{j'}^z\sigma_{j''}^z)/4,
  \end{equation}
  where for a bond $ij$, the other two neighbors of $i$ are $i'$ and $i''$, and the other two neighbors of $j$ are $j'$ and $j''$.

  For $N=20$, we see nonmonotonicity in $D$ arising from competition between $\mathcal H_D$ and $\mathcal H_I$, and this competition is reproduced by both QPUs.  Near $t_a=\SI{30}{ns}$, $D$ plateaus in the MPS data at a value of $5.45$ (consistent with $\ket{\psi_\epsilon}$), whereas QPU results tail off, thermalizing in the large-$t_a$ limit to a value of $4.8$ (consistent with $\ket{\psi_0}$).  The precise tracking of $D$ by the QPUs in the coherent region of the anneal highlights the utility of frustrated fullerenes as a quantum simulation benchmark.

In Fig.~\ref{fig:4}b, we isolate a quality measure: the fidelity with respect to the targeted perturbative ground state $\ket{\psi_\epsilon}$.  We use the Battachyyara coefficient $\mathcal F= \sum_i\sqrt{p_iq_i}$, where $p_i$ and $q_i$ are the probability weights of computational basis state $i$ in the (classical or quantum) simulation and in $\ket{\psi_\epsilon}$, respectively.  The large entropy of the ground-state manifold leads to sampling error, which we mitigate by binning probabilities based on automorphism orbits, calling the resulting ``binned fidelity'' $\mathcal F'$.  Still, sampling error remains a significant challenge in estimating $\mathcal F'$, so one must be careful to compare fidelities between equal sample sizes---in this case we use $10^5$ samples for all solvers; statistical floors are below the limits of the plots.  To effectively distinguish the quality of two simulators, sampling error must be significantly lower than the error of at least one of the simulators.

In Fig.~\ref{fig:4}b, we see a clear minimum in $1-\mathcal F'$ for the QPUs as $t_a$ increases; the increase of infidelity after this minimum is a signature of relaxation throughout the classical ground-state manifold.

\section{Discussion}

As the field of quantum simulation advances, with diverse platforms demonstrating
precise control over complex quantum states, evaluating and improving these technologies requires the development of tools that reveal nontrivial quantum behavior and facilitate direct, cross-platform comparisons between qubit
modalities.

This work introduced frustrated Ising fullerenes as a target system for quantum
simulation. We demonstrated simulation of the canonical 20-spin dodecahedron and
60-spin buckyball Ising cages. Classically, we performed MPS time-evolution
simulations for both, increasing bond dimension to 724, until the simulations no
longer fit in memory for a single GPU in the Frontier supercomputer.  (While geometry-specific tensor network methods may be effective in the short-$t_a$ regime~\cite{begusic_fast_2024,park_simulating_2025}, we found that the belief-propagation-based time-evolution method of Tindall et al.~\cite{tindall_dynamics_2025} struggles on the simulations performed here (see Fig.~\ref{fig:tns_ene}).)

Such fullerene structures can scale arbitrarily large, with numerous isomers of
equivalent size \cite{balaban_graph_1995}, enabling a wide variety of simulations
even for purely AFM systems of modest size. Simulations of such models have been
shown to be achievable on existing neutral-atom platforms~\cite{barredo_synthetic_2018}.

Precise reproduction of the inherent symmetries and degeneracies of these systems
is crucial for achieving high-quality simulation results, as they enable the
emergence of correlations from the extensive classical ground-state manifold
when quantum fluctuations are introduced. Additionally, these symmetries play a
vital role in mitigating sampling errors that can arise during Monte Carlo
inference in quantum simulators, particularly for platforms such as trapped-ion
or neutral-atom systems, where shot rates are significantly slower than those of
superconducting platforms.

In our experiments, we observed significant performance improvements across successive generations of superconducting quantum annealers, demonstrating the potential for using these systems to probe quantum phenomena and enhance emerging technologies. Ultimately, this work takes a modest but significant step toward a future scenario where classical methods are inadequate, and quantum simulators must rely on cross-platform testbeds for mutual validation.

% \bibliography{paper}

\section{Acknowledgements}
The authors thank Joel Pasvolsky for editing the manuscript.  Work was carried out under the auspices of the U.S. DOE through the Los Alamos National Laboratory, operated by Triad National Security, LLC (Contract No. 892333218NCA000001).  A.N. was supported by Natural Sciences and Engineering Research Council of Canada (NSERC) Alliance Quantum Program (Grant ALLRP-578555), CIFAR and the Canada First Research Excellence Fund, Quantum Materials and Future Technologies Program.  This research used resources of the Oak Ridge Leadership Computing Facility at the Oak Ridge National Laboratory, which is supported by the Office of Science of the U.S. Department of Energy under Contract No. DE-AC05-00OR22725.

\section{Author contributions}

ALB, KB, and AK conceived the project.  AN, JR and WB performed classical simulations.  ALB, KB and AK performed quantum simulations.  All authors contributed to the writing of the manuscript.

%\clearpage
\onecolumngrid
\vspace{8mm}\hrule\vspace{8mm}
\begin{center}
\textbf{\large Supplementary Materials:\\ \mytitle}
\end{center}\vspace{8mm}\ 
\twocolumngrid

%\tableofcontents
\setcounter{equation}{0}
\setcounter{figure}{0}
\renewcommand{\figurename}{FIG.}
\renewcommand{\thefigure}{S\arabic{figure}}
\renewcommand{\theequation}{S\arabic{equation}}
\renewcommand{\theHfigure}{S\arabic{figure}}
\renewcommand{\thefootnote}{{\Roman{footnote}}}
\makeatother

\section{Quantum annealing methods}\label{sec:qa}

\begin{figure}
  %\hrule
  \includegraphics[scale=.8]{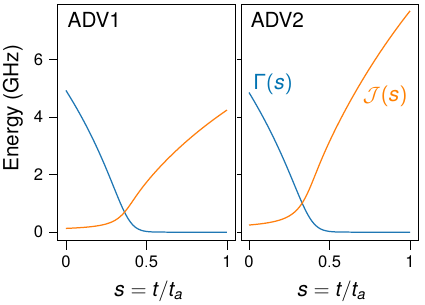}
  %\hrule
  \caption{{\bf Annealing schedules.} Schedules for the fast-anneal protocol for the Advantage (ADV1, left) and Advantage2 prototype (ADV2, right) QPUs are shown, with time-dependent transverse and Ising energy scales $\Gamma(s)$ and $\mathcal J(s)$.}\label{fig:sched}
\end{figure}

In this work we used \texttt{Advantage\_system4.1} (ADV1) and \texttt{Advantage2\_prototype2.6} (ADV2), whose annealing schedules are shown in Fig.~\ref{fig:sched}.

The fullerenes studied in this work can be embedded many times over in both QPUs; these embeddings were found using the Glasgow subgraph solver\cite{mccreesh_glasgow_2020}.  For fairness of comparison we used the same number of parallel embeddings in both chips: 49 for $N=20$, 39 for $N=24$, and 12 for $N=60$.

For each parameterization (choice of $N$, $|J_{ij}|$, $t_a$, and QPU) we iteratively shimmed per-qubit flux offsets to balance qubits at average magnetization zero, and $J_{ij}$ to balance the nearest-neighbor spin-spin correlation of couplers that are equivalent under automorphism~\cite{chern_tutorial_2023}.  These shims were run for 500 iterations, each iteration consisting of a single call to the QPU yielding 100 output samples.  The last 100 iterations were retained, yielding at least $12\times 100\times 100 > 10^5$ samples.

\section{Classical simulation methods}

\subsection{Exact diagonalization}

For $N\leq 24$ we used exact diagonalization to time-evolve the Schr\"odinger equation.  We use a uniform step in normalized time $ds = dt/t_a$, sweeping values between $ds=0.01$ and $ds=0.0001$ until satisfactory convergence of observables is reached, which occurs near $dt=\SI{0.1}{ns}$.

\subsection{Matrix product state}

\begin{figure*}
  %\hrule
  \includegraphics[scale=.8]{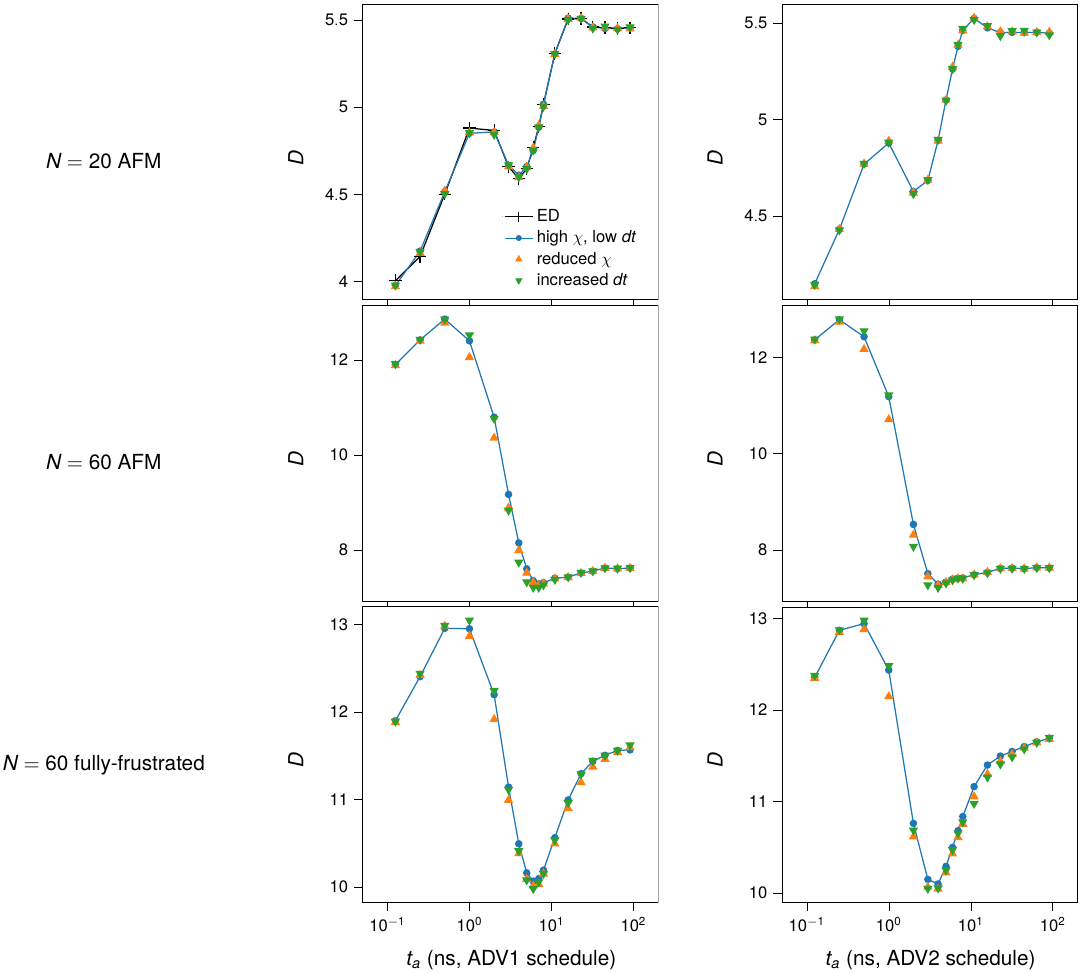}
  %\hrule
  \caption{{\bf Dimer density for various MPS parameters.} MPS experiments were run up to bond dimension $\chi=724$ and for two different $dt$ values (\ref{eq:dt}).  Blue circles show the best MPS data, for $\chi=724$, with smaller $dt$.  Orange triangles show $\chi=512$ with smaller $dt$, and green triangles show $\chi=724$ with larger $dt$.  Left and right columns indicate classical simulations using the ADV1 and ADV2 schedule, respectively.  The largest deviations appear near $t_a=\SI{1}{ns}$, arising from smaller bond dimension.  This is also where results deviate significantly from QPU results, despite being well within the QPUs' coherent regime.}\label{fig:mps_dimer}
\end{figure*}

\begin{figure*}
  %\hrule
  \includegraphics[scale=.8]{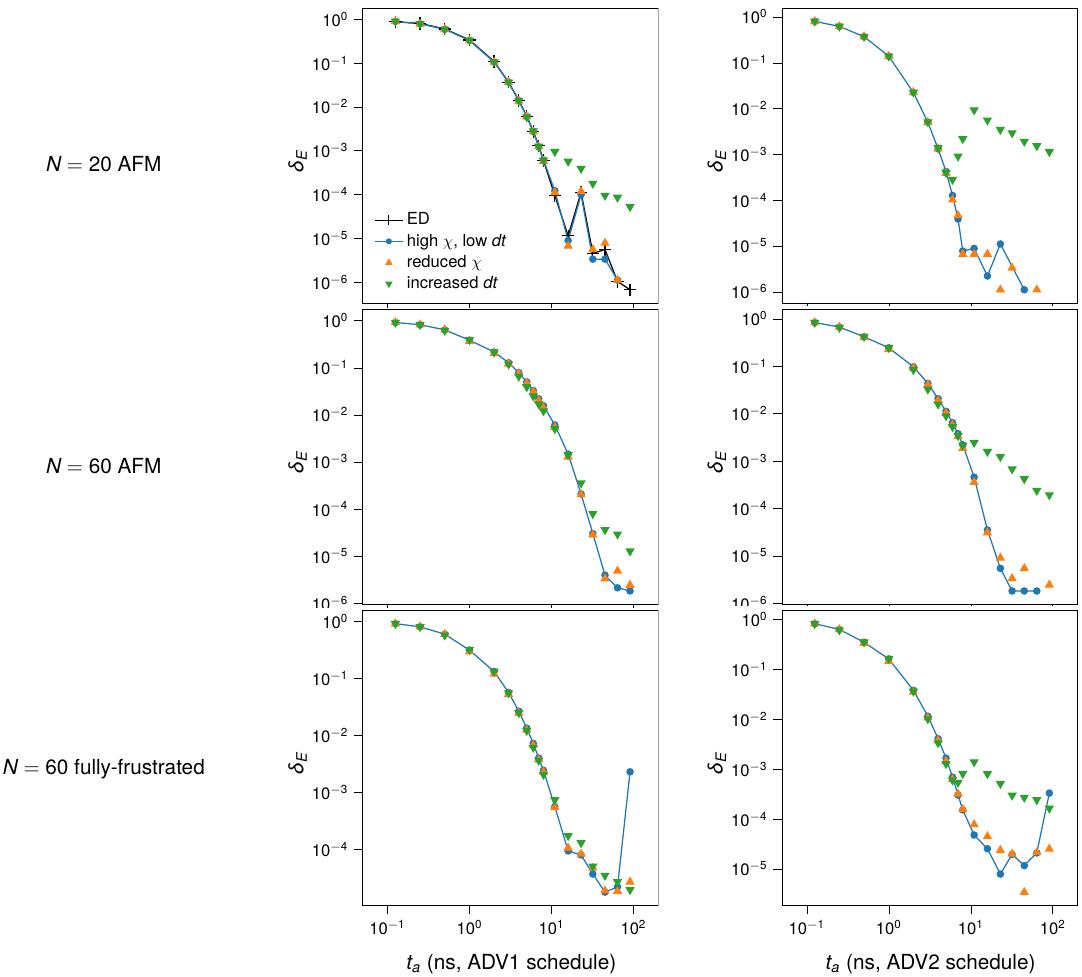}
  %\hrule
  \caption{{\bf Residual energy density for various MPS parameters.} Analogous data to Fig.~\ref{fig:mps_dimer} but for residual energy density $\delta_E$.  Large deviations are seen, particularly for long anneals when increasing the step size $dt$.}\label{fig:mps_ene}
\end{figure*}

For $N\in \{20,60\}$, i.e.~the three models presented in the main text, we ran MPS simulations with time steps
\begin{equation}\label{eq:dt}
dt=\max\{\SI{0.05}{ns},t_a/200\}\ \text{and}\ \max\{\SI{0.1}{ns},t_a/100\}
\end{equation}
and bond dimensions $\chi=512$ and $724$ on individual GPUs on the Frontier supercomputer.

For $N=20$ these bond dimensions were sufficient to converge observables to the same quality as ED (Fig.~\ref{fig:mps_dimer} for dimer density, Fig.~\ref{fig:mps_ene} for residual energy density).  For $N=60$ we see clear deviation between bond dimensions, and for longer $t_a$ in the ADV2 schedule we also see deviations between values of $dt$.

For dimer density, bond dimensions are most clearly insufficient for short anneals of $t_a\approx \SI{1}{ns}$, which is consistent with the general findings of Ref.~\onlinecite{king_beyondclassical_2025}: bond-dimension requirements for accurate simulation tend to reach a peak for simulations of this kind near this value of $t_a$.

\subsection{Tensor-network state time-evolution with belief propagation}

\begin{figure*}
  %\hrule
  \includegraphics[scale=.8]{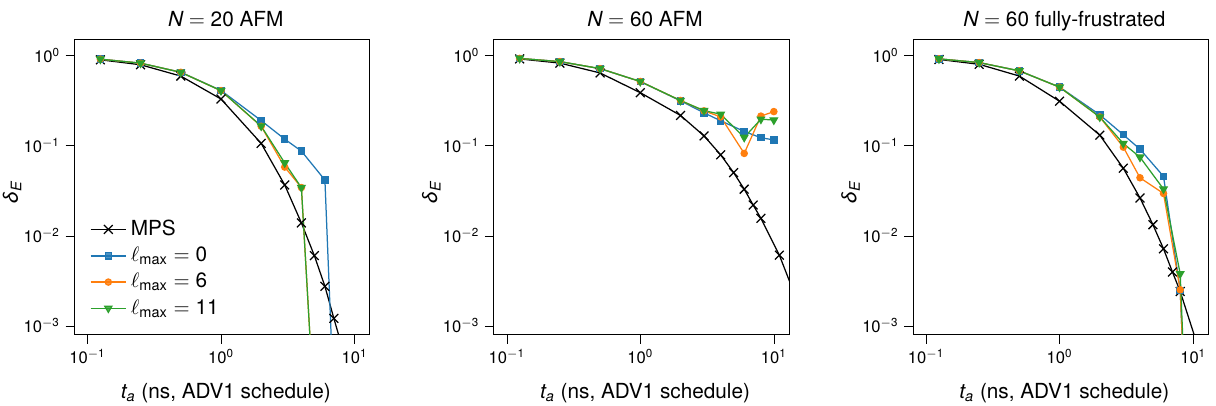}
  %\hrule
  \caption{{\bf Residual energy density for loop-corrected BP-TNS.}  Simulations use $\chi_\text{BP}=\chi=16$; MPS data shown are for $\chi=724$ with the smallest step size.  Finite loop corrections can result estimated spin-spin correlations outside the range $[-1,1]$, hence negative $\delta_E$.  Results are not sensitive to reducing $dt$.}\label{fig:tns_ene}
\end{figure*}

Lattice-specific tensor-network-state (TNS) methods using belief propagation (BP) for both time-evolution and measurement, BP-TNS, have recently shown promise in simulating quantum dynamics beyond one dimension~\cite{tindall_dynamics_2025}.  Here we run the same algorithm with the same parameters used to simulate 50-site diamond spin glasses to high accuracy: a time step $dt=\SI{0.01}{ns}$, bond dimension $\chi_\text{BP}=\chi=16$ for both time-evolution and measurement, and maximum loop-correction size $\ell_\text{max}=11$.  Existing code~\cite{tindall_code} can easily be modified to produce energies: in these fullerenes, due to symmetry the energy of a state can be determined from one ($N=20$) or two ($N=60$) nearest-neighbor correlations.  Results deviate significantly from MPS (and QA, and, for $N=20$, ED).  The more impressive performance on high-precision diamond spin glasses is likely due to the fact that the expected absolute value $|J_{ij}|$ of the weakest coupler on a loop of length $\ell$ in such a spin glass is $\tfrac{1}{\ell+1}$, i.e.,~1/7---nearly an order of magnitude weaker than in the Ising models studied here.

\section{Calibrating effective time}

\begin{figure*}
  %\hrule
  \includegraphics[scale=1]{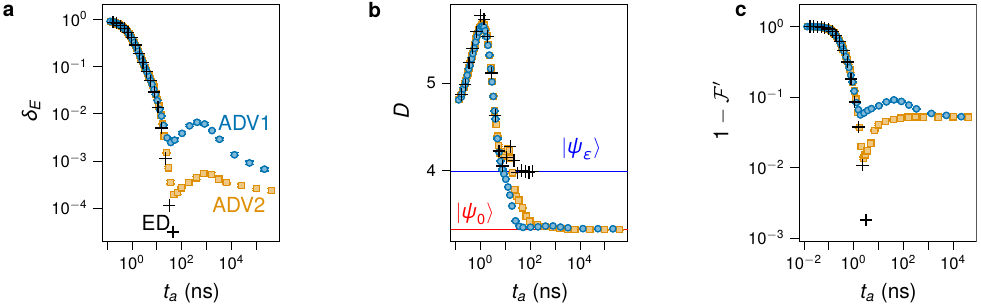}
  %\hrule
    \caption{{\bf 24-spin fullerene.}  Shown are data for the $N=24$ AFM fullerene for ADV1 and ADV2 QPUs and ED.  {\bf a}, residual energy density (as in Fig.~\ref{fig:3}.  {\bf b}, {\bf c}, dimer density and binned infidelity, as in Fig.~\ref{fig:4}.}\label{fig:N24}
\end{figure*}

QPU parameters are limited to $t_a \geq \SI{5}{ns}$ and $J_{ij}\in [-2,1]$.  Furthermore, we assume that for $t_a = \SI{5}{ns}$ the QPU dynamics closely reflect the ideal dynamics of a closed quantum system with negligible decoherence.  Under this assumption, rescaling the Hamiltonian $\mathcal H(s)$ is equivalent to rescaling time.

We cannot tune the transverse-field curve $\Gamma(s)$, but we can freely rescale couplings $J_{ij}$, thereby rescaling $\mathcal H_D$.  This is approximately equivalent to rescaling $\mathcal H(s)$ in the important region of the annealing schedule in which $\Gamma$ and $\mathcal J$ are of the same order of magnitude.

QPU runs were performed at $|J_{ij}|=0.95$ for a variety of $t_a \geq \SI{5}{ns}$ (this allows fine tuning of $J_{ij}$ terms while remaining within the permitted range $-2\leq J_{ij}\leq 1$), and for $t_a=\SI{5}{ns}$ for a variety of $|J_{ij}|\leq 0.95$.  MPS and ED runs were performed at $|J_{ij}|=1$ for a variety of $t_a$ between $\SI{0.1}{ns}$ and $\SI{128}{ns}$; MPS runs were done using both the ADV1 and the ADV2 schedule.

We mapped equivalent time between ADV1 and ADV2 MPS experiments by finding the equivalent $t_a$ in ADV2 achieving the same binned fidelity $\mathcal F'$ as $t_a=\SI{3}{ns}$ in ADV1, for $N=20$.  This gave a prefactor $\approx 1.75$ by which we multipy ADV2 time to reach equivalent ADV1 time, for all $t_a$.

 We mapped both ADV1 and ADV2 QPU experiments to ADV1 ED and MPS experiments by doing the same at $t_a=\SI{5}{ns}$, for every energy scale $|J_{ij}|$ studied, for $N=20$.  Thus QPU $t_a$ are fitted based on $\mathcal F'$ at $t_a=\SI{5}{ns}$ for $N=20$.

\section{Additional data: $N=24$}\label{sec:others}

Fig.~\ref{fig:N24} shows data for the $N=24$ antiferromagnet, for which we did not collect MPS data and which is near the limit of routine computation for exact diagonalization, taking several days on a laptop.  Results reinforce the main conclusions: both ADV1 and ADV2 provide accurate simulation for short timescales; ADV2 shows later onset of thermalization and reaches a significantly higher fidelity than ADV1 with respect to $\psi_\epsilon$.  Only ADV2 catches the small jump in $D$ near $t_a=\SI{20}{ns}$.

\end{document}